\documentclass[a4paper,11pt]{article}
\pdfoutput=1 

\usepackage{jcappub}

\usepackage{graphicx}
\usepackage{amsmath}
\usepackage{amssymb}
\usepackage{amsfonts}
\usepackage{lscape}
\usepackage{hyperref}
\usepackage{xcolor}
\usepackage{bm}
\usepackage{tabularx}
\usepackage{mathtools}
\newcommand{\be}{\begin{equation}}
\newcommand{\ee}{\end{equation}}
\newcommand{\ba}{\begin{eqnarray}}
\newcommand{\ea}{\end{eqnarray}}
\newcommand{\nn}{\nonumber}

\renewcommand{\[}{\begin{equation}}
\renewcommand{\]}{\end{equation}}
\def\be{\begin{equation}}
\def\ee{\end{equation}}
\def\bea{\begin{eqnarray}}
\def\eea{\end{eqnarray}}
\def\eqi{\begin{equation}}
\def\eqf{\end{equation}}
\def\eqia{\begin{eqnarray}}
\def\eqfa{\end{eqnarray}}
\def\lcdm{$\Lambda$CDM }

\makeatother

\title{Cosmological constraints with the Effective Fluid approach for Modified Gravity}

\author[a,1]{Wilmar Cardona,\note{Corresponding author.}}
\author[b]{Rub\'{e}n Arjona,}
\author[c]{Alejandro Estrada,}
\author[b]{Savvas Nesseris}

\affiliation[a]{Departamento de F\'isica, Universidad del Valle, Ciudad Universitaria Mel\'endez, 760032, Cali, Colombia}

\affiliation[b]{Instituto de F\'isica Te\'orica UAM-CSIC, Universidad Auton\'oma de Madrid,
Cantoblanco, 28049 Madrid, Spain}

\affiliation[c]{Departamento de Matem\'aticas, Universidad del Valle, Ciudad Universitaria Mel\'endez, 760032, Cali, Colombia}

\emailAdd{wilmar.cardona@correounivalle.edu.co}
\emailAdd{ruben.arjona@uam.es}
\emailAdd{alejandro.estrada@correounivalle.edu.co}
\emailAdd{savvas.nesseris@csic.es}

\date{\today}

\abstract{
Cosmological constraints of Modified Gravity (MG) models are seldom carried out rigorously. First, even though general MG models evolve differently (i.e., background and perturbations) to the standard cosmological model, it is usual to assume a $\Lambda$CDM background. This treatment is not correct and in the era of precision cosmology could induce undesired biases in cosmological parameters. Second, neutrino mass is usually held fixed in the analyses which could obscure its relation to MG parameters. In a couple of previous papers we showed that by using the Effective Fluid Approach we can accurately compute observables in fairly general MG models. An appealing advantage of our approach is that it allows a pretty easy implementation of this kinds of models in Boltzmann solvers (i.e., less error--prone) while having a useful analytical description of the effective fluid to understand the underlying physics. This paper illustrates how an effective fluid approach can be used to carry out proper analyses of cosmological constraints in MG models. We investigated three MG models including the sum of neutrino masses as a varying parameter in our Markov Chain Monte Carlo analyses. Two models (i.e., Designer $f(R)$ [DES-fR] and Designer Horndeski [HDES]) have a background matching $\Lambda$CDM, while in a third model (i.e., Hu $\&$ Sawicki $f(R)$ model [HS]) the background differs from the standard model. In this way we estimate how relevant the background is when constraining MG parameters along with neutrinos' masses. We implement the models in the popular Boltzmann solver CLASS and use recent, available data (i.e., Planck 2018, CMB lensing, BAO, SNIa Pantheon compilation, $H_0$ from SHOES, and  RSD Gold-18 compilation) to compute tight cosmological constraints in the MG parameters that account for  deviation from the $\Lambda$CDM model. For both the DES-fR and the HS model we obtain $\log_{10}b<-8$ at $68\%$ confidence when all data are included. In the case of the HDES model we find a somewhat weaker value of $\log_{10}J_c>-5$ at $68\%$ confidence. We also find that constraints on MG parameters are a bit weakened when compared to the case where neutrinos' masses are held fixed in the analysis. 
}
\begin{document}
\maketitle

\section{Introduction \label{Section:Introduction}}

The growing evidence for the late-time accelerating expansion of the Universe represents a major milestone in cosmology \cite{Riess:1998cb,Perlmutter:1998np,Aghanim:2018eyx,Abbott:2018wzc} and investigations using Machine Learning algorithms have confirmed this fact through model independent analyses \cite{Arjona:2019fwb,Bengaly:2019ibu,Arjona:2020kco}. Bayesian analyses indicate that the standard model of cosmology \lcdm is preferred over its alternatives due to its simplicity and lower number of free parameters \cite{Heavens:2017hkr}. The concordance model is relatively simple and having just six free parameters is able to make predictions which agree remarkably well with most observations \cite{Aghanim:2018eyx,Abbott:2018wzc,To:2020bhf,Mossa:2020gjc}. Nevertheless, the cosmological constant problem, our current ignorance on the nature of Dark Matter (DM) as well as a worrying discordance in a few cosmological parameters represent big disadvantages for the \lcdm model.

Over time the disagreement between the Hubble constant determined via distance ladder and the $H_0$ value obtained through analyses of the CMB has become more interesting \cite{DiValentino:2020zio}. Although the discrepancy could be due to unaccounted-for systematic errors, there also exists the more appealing possibility of new physics (e.g.  Modified Gravity, early Dark Energy). With the coming of latest data analyses disagreements on the values of $\Omega_m$ and $\sigma_8$ also became apparent \cite{DiValentino:2020vvd} or for example, a $\sim4\sigma$ deviation of the dark energy equation of state $w(z)$ from the \lcdm model using quasars at high redshift up to $z\sim7.5$ \cite{Risaliti:2018reu}.
The curvature of the Universe has also given rise to a discussion on recent analyses \cite{DiValentino:2019qzk,Aghanim:2018eyx,Handley:2019tkm,Efstathiou:2020wem,Vagnozzi:2020zrh,Vagnozzi:2020dfn}. There are some mild hints ($\approx 2\sigma$) of Modified Gravity (MG)  \cite{Aghanim:2018eyx,Ade:2015rim}, which challenge assumptions made in the standard model. Here we will investigate some viable MG models in light of latest data releases.

Dark Energy (DE) and MG have emerged as alternatives to the cosmological constant model. In the case of DE, some of the simplest models include minimally coupled scalar fields in the form of quintessence, that has a standard kinetic term, or k-essence which has a generalized kinetic-term \cite{Copeland:2006wr}. On the other hand, MG models are covariant modifications to General Relativity (GR) that extend the Einstein-Hilbert action in various ways: either by promoting it to a function as in the $f(R)$ models and by introducing higher order curvature invariants (see for example \cite{Clifton:2011jh}) or by adding non-minimally coupled scalar fields, as in the case of Horndeski theory \cite{Horndeski:1974wa}. MG models have the advantage that they are also inspired from high-energy physics, as covariant modifications to GR of a similar form appear naturally when one tries to renormalize GR at one loop order \cite{Birrell:1982ix}.

The advantages of these alternatives, especially of the MG models, are clear: there is no need for a cosmological constant as the accelerating expansion of the Universe can be explained by the presence of the extra terms in the modified Friedmann equations due to the new degrees of freedom. Both DE and MG models are also able to describe well the cosmological observations and give equally good fits to the data as the \lcdm model. 

On the other hand, there are also some disadvantages for these models, for example the presence of the additional parameters may penalize some of the models when one calculates the Bayesian evidence and uses the Jeffreys' scale, albeit it has been shown that the latter has to be interpreted with care \cite{Nesseris:2012cq}. Furthermore, by using N-body simulations it has been shown that a compelling $f(R)$ model fails to reproduce the observed redshift-space clustering on scales $\sim 1-10 ~ \mathrm{Mpc}~ h^{-1}$ \cite{He:2018oai}. Finally, as both the extra scalar field degrees of freedom and the higher order corrections to GR are as yet unobserved in a laboratory or in an astrophysical setting, their motivation is obviously somewhat weakened. Not only that, but recently several MG models of the Horndeski type have been ruled out, via the measurement of the speed of propagation of the gravitational waves by the event GW170817 and its optical counterpart GRB170817A \cite{Monitor:2017mdv,Ezquiaga:2017ekz,Creminelli:2017sry}. Thus, as the available parameter space has shrank remarkably, there are a few remaining models which deserve attention as well as proper analyses. 

However, many analyses of the remaining models, especially the ones where the background expansion differs significantly from the \lcdm model, do not consider the background expansion properly and just fix it to either the \lcdm or a constant $w$ model, as was observed in Refs.~\cite{Arjona:2018jhh, Arjona:2019rfn}. This obviously biases the results as it introduces biases in the cosmological parameters and spurious tensions with the data. However, some recent analyses have also acknowledged this discrepancy and newer versions of the Boltzmann solvers now have support for the correct backgrounds in some cases \cite{Zucca:2019xhg}. 

On the other hand, the so-called Effective Fluid Approach has the advantage that it presents a unified approach to analyse all models under the same umbrella, allows for the correct background expansion in the models, all without sacrificing the accuracy of the results \cite{Arjona:2018jhh, Arjona:2019rfn,Arjona:2020gtm}. In a nutshell, the Effective Fluid Approach works by rewriting the field equations of the MG model as GR and a DE fluid with an equation of state $w$, a pressure perturbation $\delta P$ and an anisotropic stress $\sigma$. Especially the latter is crucial as sometimes it is ignored in analyses of MG models \cite{Sawicki:2015zya}, something which might bias the results \cite{Cardona:2019qaz}. Moreover, through a joint Machine Learning analysis applied to the latest cosmological data hints of dark energy anisotropic stress were found \cite{Arjona:2020kco}. 

In the Effective Fluid Approach we also assume that in the relevant scales, where linear theory applies, the sub-horizon and quasi-static approximations hold. With these, general analytical expressions for the equation of state, the pressure perturbation, and the anisotropic stress were found in Refs.~\cite{Arjona:2018jhh, Arjona:2019rfn}. With the latter, one may then just solve numerically the evolution equations for the perturbations, found for example in Ref.~\cite{Ma:1995ey}. 

Thus, the main advantage of the Effective fluid approach is that once one has the expressions for the variables $w$, $\delta P$ and $\sigma$, it is very straightforward to also implement them in standard Boltzmann codes, such as CLASS, with very minimal modifications. In fact, in \cite{Arjona:2018jhh, Arjona:2019rfn}, this was done with the EFCLASS code, which implements the aforementioned approach, where it was found that EFCLASS and hi-CLASS \cite{Zumalacarregui:2016pph}, a modification of CLASS that solves numerically the whole set of perturbation equations for Horndeski, agree to better than $0.1\%$ \cite{Arjona:2019rfn}.

Recently, a comparison of different approaches to the quasi-static approximation in Horndeski models was made by Ref.~\cite{Pace:2020qpj}, by applying this approximation to either the field equations, as done in the Effective fluid approach, or the equations for the two metric potentials $\Phi$ and $\Psi$ and finally,  the use of the attractor solution derived within the Equation of State approach \cite{Battye:2015hza}. It was found that all three approaches agree exactly on small scales and that in general, this approach is valuable in future model selections analyses for models beyond the \lcdm model.

In this analysis we use the Effective Fluid approach and our EFCLASS code, for the background and first order perturbations, to obtain cosmological constraints with the latest cosmological data sets: we include the Pantheon SNe compilation \cite{Scolnic:2017caz}, the Planck 2018 CMB data \cite{Aghanim:2018eyx}, the $H_0$ Riess measurement \cite{Riess:2019cxk}, various BAO points \cite{Alam:2016hwk,Beutler:2011hx,Ross:2014qpa}, and a new redshift space distortions (RSD) likelihood (see Ref.~\cite{Arjona:2020yum} for the ``Gold 2018" compilation of Ref.~\cite{Sagredo:2018ahx}). An important aspect of our investigation is that we take into consideration neutrino mass as a varying parameter. Neutrino mass is usually held fixed in analyses which could obscure its relation to MG parameters. The role of massive neutrinos in modified gravity was first investigated in \cite{Motohashi:2010sj} by considering $f(R)$ gravity. The implementation in Boltzmann solvers of $f(R)$ gravity including neutrino mass as a varying parameter was carried out in \cite{Motohashi:2012wc,Chudaykin:2014oia}, where cosmological constraints were also computed.   

The paper is organised as follows. In Sec.~\ref{Section:models} we briefly summarize our theoretical framework and present the models we consider, in Sec.~\ref{Section:cosmo-constraints} we present the results of our MCMC analysis with EFCLASS, in Sec.~\ref{Section:conclusions} we conclude  and lastly, in Appendix \ref{Section:appendix-A} we present some details on our RSD likelihood. 

\section{Theoretical framework \label{Section:models}}

The standard cosmological model \lcdm assumes the Einstein-Hilbert action
\be 
\mathcal{L} = \frac{1}{2\kappa}f(R) + \mathcal{L}_m
\label{Eq:Einstein-Hilbert-action}
\ee
with $f(R)=R$, $R$ the Ricci scalar, $\mathcal{L}_m$ the Lagrangian for matter fields, and the constant $\kappa \equiv 8\pi G_N$ with $G_N$ being the bare Newton's constant. We can derive the field equations by applying the Principle of Least Action; they read
\be 
\label{EinsteinEq}
G_{\mu\nu} = \kappa T^{(m)}_{\mu\nu}
\ee
where $G_{\mu\nu}\equiv R_{\mu\nu}-\frac{1}{2}g_{\mu\nu}R$ is the Einstein tensor, $R_{\mu\nu}$ is the Ricci tensor, $g_{\mu\nu}$ is the metric, and $T^{(m)}_{\mu\nu}$ is the energy-momentum tensor of matter fields.\footnote{Throughout this paper our conventions are: $(-+++)$ for the metric signature, the Riemann and Ricci tensors are given respectively by $V_{b;cd}-V_{b;dc}=V_{a}R^{a}_{\;bcd}$ and $R_{ab}=R^{s}_{\;asb}$.} It is also possible to consider more general theories (e.g., $f(R)\neq R$, Horndeski theories) and derive similar field equations. As we clearly explained in Refs. \cite{Arjona:2018jhh,Arjona:2019rfn}, modifications to GR in these kinds of MG models can be interpreted as an effective fluid yielding field equations which schematically look like
\bea
G_{\mu\nu}&=&\kappa\left(T_{\mu\nu}^{(m)}+T_{\mu\nu}^{(\textrm{DE})}\right).
\label{eq:effEqs}
\eea
Here the tensor $T_{\mu\nu}^{(\textrm{DE})}$ depends on the metric and its derivatives, and in the case of general scalar-tensor theories, also on the scalar field and its derivatives (see \cite{Arjona:2019rfn}).

Observational evidence indicates that the Universe is statistically homogeneous and isotropic on large scales \cite{Hogg:2004vw,PhysRevD.74.123521,PhysRevLett.107.041301,Akrami:2019bkn}. Therefore we will as usual assume a flat linearly perturbed Friedmann-Lema\^{i}tre-Robertson-Walker (FLRW) metric
\be
ds^2=a(\tau)^2\left[-(1+2\Psi(\vec{x},\tau))d\tau^2+(1-2\Phi(\vec{x},\tau))d\vec{x}^2\right],
\label{eq:FRWpert}
\ee
where $a$ is the scale factor, $\vec{x}$ represents spatial coordinates, $\tau$ is the conformal time, and $\Psi$ and $\Phi$ are the gravitational potentials in the Newtonian gauge. 

In order to describe matter fields, we will consider them as ideal fluids having small perturbations. We will take into consideration an Effective Fluid Approach (EFA) when dealing with MG models: modifications to GR will be described by an effective fluid having equation of state, pressure perturbation, and anisotropic stress. In our approach, and assuming a metric \eqref{eq:FRWpert}, the background evolution is governed by the Friedmann equations
\bea
\mathcal{H}^2&=&\frac{\kappa}{3}a^2 \left(\bar{\rho}_{m}+\bar{\rho}_{\textrm{DE}}\right), \\
\dot{\mathcal{H}}&=&-\frac{\kappa}{6}a^2 \left(\left(\bar{\rho}_{m}+3\bar{P}_{m}\right)+\left(\bar{\rho}_{\textrm{DE}}+3\bar{P}_{\textrm{DE}}\right)\right),
\eea
where $\mathcal{H}\equiv\dot{a}/a$ is the conformal Hubble parameter.\footnote{In our notation a dot over a function denote derivative respect to the conformal time $\dot{f}=df/d\tau$. The Hubble parameter $H$ and the conformal Hubble parameter $H$ are related via $\mathcal{H}=aH$.} Here $\bar{\rho}_{\textrm{DE}}$  and $\bar{P}_{\textrm{DE}}$ respectively denote density and pressure of the DE effective fluid. The effective DE equation of state $w_{\textrm{DE}} \equiv \bar{P}_{\textrm{DE}}/\bar{\rho}_{\textrm{DE}}$ allows us to describe the background evolution in these kinds of models.
Another ingredient that we need is the evolution equations for the perturbations obtained through the energy-momentum conservation $T^{\mu\nu}_{;\nu}=0$ \cite{Arjona:2018jhh}:
\bea
\delta' &=& 3(1+w) \Phi'-\frac{V}{a^2 H}-\frac{3}{a}\left(\frac{\delta P}{\bar{\rho}}-w\delta\right),
\label{eq:evolution-delta}
\eea
\bea
V' &=& -(1-3w)\frac{V}{a}+\frac{k^2}{a^2 H}\frac{\delta P}{\bar{\rho}} +(1+w)\frac{k^2}{a^2 H} \Psi -\frac23 \frac{k^2}{a^2 H} \pi,
\label{eq:evolution-V}
\eea
where $V$ is the scalar velocity perturbation, $\delta P$ is the pressure perturbation, $\pi$ is related to the anisotropic stress as $\pi \equiv \frac{3}{2}(1+w)\sigma$ and the prime $'$ is the derivative with respect to the scale factor $a$. In Ref.~\cite{Arjona:2019rfn} we extensively used the EFA on the remaining Horndeski Lagrangian. By applying both sub-horizon and quasi-static approximations, we managed to find analytical expressions for several quantities describing Horndeski models as an effective DE fluid, namely, DE pressure perturbation $\frac{\delta P_{\textrm{DE}}}{\overline{\rho}_{DE}}(a,k) $, DE scalar velocity perturbation $V_{\textrm{DE}}(a,k)$, and DE anisotropic stress $\pi_{\textrm{DE}}(a,k)$.

However, note that when one tries to mimic the \lcdm expansion in certain classes of theories, e.g. DHOST, instabilities may appear in the perturbations as the sound speed becomes negative and the so-called ``scordatura" correction is necessary to balance the equations. In our case we have already checked that our $f(R)$ model does not suffer from this issue as even though the sound speed is negative, it is balanced by the anisotropic stress such that the ``effective sound speed”, is positive and the perturbations are stable, see Eq. (48) and Fig. 2 in Ref.~\cite{Arjona:2019rfn}.  In the case of the HDES designer model we require that at all times there be no ghostly instabilities and that GR be recovered at early times (see Sec.~V in Ref.~\cite{Arjona:2019fwb}). 

Concerning strongly coupled issues when taking into consideration the sub-horizon and quasi-static approximations we note the following. In the case of the $f(R)$ models in Ref.~\cite{Arjona:2018jhh} we found analytical expressions for the effective Newton constant which are well-defined under these approximations. For the HDES designer model we also found a well-defined effective Newton's constant (see for instance Eq.~(177) in Ref.~\cite{Arjona:2019fwb}). For the particular models used in our analysis, we have already shown that the perturbations are stable by comparing our solutions with the full numerical solutions, see Figs. 3 and 4 in Ref.~\cite{Arjona:2019rfn} and Fig. 5 in Ref. \cite{Arjona:2019fwb}. So, while the scordatura correction is necessary for the quasi-static limit in some cases, here we did not need to take it into account. 

Having mapped remaining Horndeski models into an effective DE fluid, one can easily implement them in Boltzmann codes which compute observables such as the CMB angular power spectrum and the matter power spectrum. In the following subsections we will provide a few details on the specific models that we study in this work.

\subsection{DES-fR model}

The remaining Horndeski Lagrangian includes a very important sort of MG models, namely, $f(R)$. In \cite{Arjona:2018jhh} we showed there is a correspondence between a given $f(R)$ model and its effective DE equation of state $w_{DE}$. Then it becomes clear that by specifying a background (i.e., an equation of state), it is in principle possible to obtain a corresponding $f(R)$. The models obtained in this way are the so-called designer f(R) models \cite{Multamaki:2005zs,delaCruzDombriz:2006fj,Nesseris:2013fca}. An interesting case is the $w_{DE}=-1$ designer $f(R)$ model (DES-fR, henceforth) that mimics the standard \lcdm model at the background, while exhibiting differences in the evolution of the linear perturbations.



The DES-fR model satisfying all viability conditions (see, for instance, \cite{Pogosian:2007sw}) is given by \cite{Nesseris:2013fca}
\bea
\label{des}
f(R)&=&R-2\Lambda+\alpha~H_0^2\left(\frac{\Lambda }{R-3 \Lambda }\right)^{c_{0}} \times  {}_2F_1\left(c_{0},\frac{3}{2}+c_{0},\frac{13}{6}+2c_{0},\frac{\Lambda }{R-3 \Lambda }\right)\;,
\eea
where $c_{0}=\frac{1}{12} \left(-7+\sqrt{73}\right)$, $\alpha$ is a free dimensionless parameter, $H_0$ is the Hubble constant, $\Lambda$ is a constant, and ${}_2F_1$ is a hypergeometric function. In the literature it is common to define $F\equiv f'(R)$ and we will follow this convention. Instead of using $\alpha$ in Eq.  \eqref{des} we can parametrise our expressions in terms of $b_\pi = f_{R,0} \equiv F(a=1)-1$. Furthermore, in Ref.~\cite{Arjona:2018jhh} we found that in the range $a\in[0,1]$ the following approximation around $a\simeq 0$ is equally accurate
\bea
F(a)\simeq 1 + f_{R,0}\frac{\Omega_{m0}^{-c_0-1}}{\, _2F_1\left(c_0+1,c_0+\frac{3}{2};2 c_0+\frac{13}{6};1-\Omega_{m0}\right)}a^{3 (1 + c_0)} +\mathcal{O}(a^{3(2+c_0)}) , \label{eq:designF}
\eea
Then, when we rewrite our expressions for the DE effective fluid in terms of $f_{R,0}$ we find that they depend on 
\be
g(x) = {}_2F_1(\frac{3}{2}+c_0,2+c_0;2c_0+\frac{13}{6},x),
\label{Eq:g-of-x}
\ee
where
\be 
x \equiv \frac{a^3(\Omega_{m,0}-1)}{a^3(\Omega_{m,0}-1)-\Omega_{m,0}}.
\ee 

Implementing special functions in EFCLASS is not an easy task. Therefore, we used a Taylor approximation for $g(x)$ in Eq.~\eqref{Eq:g-of-x} which works really well around $x=0$ while keeping $30$ terms in the expansion. 

\subsection{HS model}

The popular Hu $\&$ Sawicki model\footnote{The Starobinsky model \cite{Starobinsky:2007hu} has a pretty similar $f(R)=R-c_1~m^2 \left[1-\left(1+R^2/m^{4}\right)^{-n}\right]$ and was independently and almost simultaneously introduced in the literature. As we pointed out in \cite{Arjona:2018jhh}, since the results we obtain for the HS model are very similar to those for the Starobinsky model, in order to keep our presentation simple we only present results for the HS model.} (HS, henceforward) \cite{Hu:2007nk}  
\begin{equation}
\label{Hu}
f(R)=R-m^2 \frac{c_1 (R/m^2)^n}{1+c_2 (R/m^2)^n},
\end{equation}
can actually be rewritten, after some algebraic manipulations, as \cite{Basilakos:2013nfa}
\be
\label{Hu1}
f(R)= R- \frac{2\Lambda }{1+\left(\frac{b_{hs} \Lambda }{R}\right)^n},
\ee
where $\Lambda= \frac{m^2 c_1}{2c_2}$ and $b_{hs}=\frac{2 c_2^{1-1/n}}{c_1}$. In \cite{Basilakos:2013nfa} the authors found that when written in the form \eqref{Hu1}, it is clear the reason why the HS model satisfies solar system tests: if $b_{hs} \to 0$ \lcdm is recovered, and  if $b_{hs} \to \infty$ a matter dominated universe is obtained  i.e.,
\bea
 \lim_{b_{hs}\rightarrow0}f(R)&=&R-2\Lambda , \nn \\
\lim_{b_{hs}\rightarrow \infty}f(R)&=&R.
\eea
If the parameter $b_{hs}$ is small enough, the HS model can be regarded as a ``perturbation'' around \lcdm. This is key in our work since we do not assume the usual approximation of fixing the background to \lcdm when investigating the HS model. However, solving numerically the equation for the Hubble parameter is not trivial in the HS model and we worked out a different approach. Instead of approximating the background, we solve the field equations and find an approximate, accurate analytical expression for the Hubble parameter. We followed the treatment in \cite{Basilakos:2013nfa}, where it was shown that HS Hubble parameter can be written as
\bea
\label{HSHubbleApproximation}
H_{HS}(a)^2 = H_{\Lambda}(a)^2 + b_{hs} \, \delta H_1(a)^2 + b_{hs}^2\, \delta H_2(a)^2 + \hdots,~~
\eea
which is an analytical approximation that works extremely well, for example for $b_{hs}\leq 0.1$ the average error with respect to the numerical solution is $10^{-5}\%$ for redshifts $z\leq 30$.

We had already shown in Ref.~\cite{Arjona:2018jhh} that neglecting the correct treatment of the background for the HS model may lead to theoretical biases and incorrect estimations of the CMB power spectrum. This is particularly clear in Fig. 7 of Ref.~\cite{Arjona:2018jhh}, where we compared the temperature CMB spectra of the HS model using three different Boltzmann codes: our EFCLASS, MGCAMB and FRCAMB, of which only ours (at the time) treated correctly the background. As can be seen, the difference in the spectra is on the order of $10\%$ at $l<5$. Thus, we feel that including the proper background evolution for the HS model is necessary.

Of course, one would expect that using the full data, as we will also see in later sections, the best-fit value of $b_{hs}$ will be expected to be small. However, we would like to stress that this cannot be used a posteriori to justify neglecting the correct background expansion history, as it might lead to a ``vicious circle": we assume the background is given by $\Lambda$CDM, when in fact it is that of the HS $f(R)$ model, fit the data and find consistency with $\Lambda$CDM, thus use that to justify neglecting the correction in the first place. In any case, using the correct expressions in this case is trivial, as show in Ref.~\cite{Arjona:2018jhh}.

\subsection{HDES model}
In what follows, we will present a brief overview of a family of designer Horndeski (HDES) models that have been already studied in Ref.~\cite{Arjona:2019rfn}. These are models whose background is exactly that of the \lcdm model but at the perturbation level it is dictated by the Horndeski theory. In its full form, Horndeski theory constitutes as the most general Lorentz-invariant extension of GR in four dimensions and contains a few DE and MG models. Due to the recent discovery of gravitational waves by the LIGO Collaboration the Horndeski Lagrangian has been severely reduced. In particular, it has been found the following constrain on the speed of GWs \cite{Ezquiaga:2017ekz}
\bea
-3 \cdot 10^{-15} \le c_g/c-1 \le 7 \cdot 10^{-16},
\eea
which implies that for Horndeski theories
\bea
G_{4X}\approx 0, \hspace{2mm} G_5 \approx \text{constant}.
\eea
Then, the surviving part of the Horndeski Lagrangian reads,
\be
S[g_{\mu \nu}, \phi] = \int d^{4}x\sqrt{-g}\left[\sum^{4}_{i=2} \mathcal{L}_i\left[g_{\mu \nu},\phi\right] + \mathcal{L}_m \right],
\label{eq:action1}
\ee
where
\bea
\mathcal{L}_2&=& G_2\left(\phi,X\right) \equiv K\left(\phi,X\right),\\
\mathcal{L}_3&=&-G_3\left(\phi,X\right)\Box \phi,\\
\mathcal{L}_4&=&G_4\left(\phi\right) R,
\eea
and $\phi$ is a scalar field, $X \equiv -\frac{1}{2}\partial_{\mu}\phi\partial^{\mu}\phi$ is a kinetic term, and $\Box \phi \equiv g^{\mu \nu}\nabla_{\mu}\nabla_{\nu}\phi$; $K$, $G_3$ and $G_4$ are free functions of $\phi$ and $X$. From the action \eqref{eq:action1} one can find several theories, for example $f(R)$ theories \cite{Chiba:2003ir}, Brans-Dicke theories, \cite{Brans:1961sx} and Cubic Galileon \cite{Quiros:2019ktw}. The HDES family of models that where constructed in Ref.~\cite{Arjona:2019rfn} limit to the Kinetic Gravity Braiding (KGB) which is distinguished by the following functions
\begin{equation}
K=K(X), \hspace{5mm} G_3=G_3(X), \hspace{5mm} G_4=\frac{1}{2 \kappa}.\label{eq:KGB:definition}
\end{equation}
With Eq.~\eqref{eq:KGB:definition}, we will present how to find a specific family of designer models such that $w_{\textrm{DE}}=-1$, i.e., the background is always that of the \lcdm model but at the perturbation level it follows Horndeski's theory. The usefulness of this designer model comes from allowing one to detect deviations from \lcdm at the perturbations level and is a natural expansion of our previous work \cite{Nesseris:2013fca,Arjona:2018jhh}. For our HDES model we need two functions, the modified Friedmann equation and the scalar field conservation equation, both of which have been presented and analyzed in detail in \cite{Arjona:2019rfn}. The modified Friedmann equation reads, 
\bea
\label{eq:friedeq}
 &-H(a)^2-\frac{K(X)}{3}+H^2_0\Omega_m(a)+\nn\\
 &+2\sqrt{2}X^{3/2}H(a)G_{3X}+\frac{2}{3}X K_X=0,
\eea
where $\Omega_m(a)$ is the matter density and $H_0$ is the Hubble parameter. The scalar field conservation equation is
\begin{equation}
\label{eq:scfeq}
    \frac{J_c}{a^3}-6XH(a)G_{3X}-\sqrt{2}\sqrt{X}K_X=0.
\end{equation}
The constant $J_c$ quantifies our deviation from the attractor given the KGB model \cite{Kimura:2010di}. By looking at Eqs.~\eqref{eq:friedeq} and \eqref{eq:scfeq} we see that we have three unknown functions $G_{3X}(X)$, $K(X)$ and $H(a)$, hence the system is undetermined. Then, we need to describe one of the three unknown functions $G_{3X}(X)$, $K(X)$ and $H(a)$, and find out the other two using Eqs.~\eqref{eq:friedeq} and \eqref{eq:scfeq}. For convenience, we write the Hubble parameter as a function of the kinetic term $X$, i.e., $H=H(X)$ and then solve the previous equations to find $(G_{3X}(X),K(X))$. Then we find
\bea
\label{eq:systemdes}
K(X) &=& -3 H_0^2 \Omega_{\Lambda,0}+\frac{J_c \sqrt{2X} H(X)^2}{H_0^2 \Omega_{m,0}}-\frac{J_c \sqrt{2X} \Omega_{\Lambda,0}}{\Omega_{m,0}}, \nn\\
G_{3X}(X) &=& -\frac{2 J_c H'(X)}{3 H_0^2 \Omega_{m,0}},
\eea
where $\Omega_{m,0}$ is the matter density at redshift $z=0$ and $\Omega_{\Lambda,0}=1-\Omega_{m,0}$. With Eqs.~\eqref{eq:systemdes} we can create a whole family of designer models. A specific model found, dubbed HDES \cite{Arjona:2019rfn}, that has a smooth limit to \lcdm and also recovers GR when $J_c\sim0$ is the following. First, we demand that the kinetic term behaves as $X= \frac{c_0}{H(a)^n}$, where $c_0>0$ and $n>0$. Then, from Eqs.~\eqref{eq:scfeq} and \eqref{eq:friedeq} we have
\bea
\label{eq:bestdes}
G_{3}(X)&=&-\frac{2 J_c c_0^{1/n} X^{-1/n}}{3 H_0^2 \Omega_{m,0}},\\
K(X)&=&\frac{\sqrt{2} J_c c_0^{2/n} X^{\frac{1}{2}-\frac{2}{n}}}{H_0^2 \Omega_{m,0}}-3 H_0^2 \Omega_{\Lambda,0}-\frac{\sqrt{2} J_c \sqrt{X} \Omega_{\Lambda,0}}{\Omega_{m,0}}.\nn
\eea
A particular HDES model that we have selected for our MCMC runs and we also used it in our comparison with hi-CLASS previously \cite{Arjona:2019rfn} comes by setting $n=1$ in Eq.~(\ref{eq:bestdes}). In the effective fluid approach one needs to specify three functions: the equation of state, the pressure perturbation and the anisotropic stress $(w,\delta P,\sigma)$ to modify properly the background and the linear perturbations in the CLASS code. With the HDES model we only need to specify $\delta P$ since $w=-1$ and $\sigma=0$ for this particular model. As it was shown in \cite{Arjona:2019rfn}, computing $\delta P$ through the effective fluid approach and then introducing this function into the evolution equation for the perturbations Eqs.~(\ref{eq:evolution-delta}) and (\ref{eq:evolution-V}) we can specify the DE scalar velocity perturbation $V_{\textrm{DE}}$ as
\be
V_{\textrm{DE}}\simeq \left(-\frac{14 \sqrt{2}}{3} \Omega_{m,0}^{-3/4} J_c~ a^{1/4}\right)\frac{\bar{\rho}_m}{\bar{\rho}_{\textrm{DE}}} \delta_m.\label{eq:VDEHDES}
\ee
which is what we implement later in our modified version of CLASS. 

\section{Cosmological Constraints \label{Section:cosmo-constraints}}

\subsection{Methodology}

In order to compute cosmological constraints we use the following data sets. Firstly, we utilise the 2018 release by the Planck Collaboration including temperature and polarisation anisotropies of the CMB (\texttt{TTTEEE}) as well as CMB lensing (\texttt{lensing}) \cite{Aghanim:2018eyx}. Secondly, we include measurements of Baryon Acoustic Oscillations (\texttt{BAO}) from Refs. \cite{Alam:2016hwk,Beutler:2011hx,Ross:2014qpa}. Thirdly, Pantheon supernovae (\texttt{SNe}) from \cite{Scolnic:2017caz} were also incorporated in the analysis. Fourthly, we employed local Hubble measurement (\texttt{H0}) from Ref. \cite{Riess:2019cxk} as a Gaussian prior. Finally, we coded a new likelihood for a compilation of Redshift-Space-Distortions (\texttt{RSD}) measurements (see Appendix \ref{Section:appendix-A} and Ref. \cite{Arjona:2020yum}). 

\begin{table}[!t]
\centering
\begin{tabular}{@{}cc}
\hline
Parameter & Range \\
\hline
$ \log_{10} b_\pi$ & $[-10,0]$\\
$ \log_{10} b_{hs}$ & $[-10,-1]$\\
$ \log_{10} J_c$ & $[-10,0]$\\
\hline
\end{tabular}
\caption{Flat prior bounds used in the MCMC analyses. Prior range for other parameters is set as in Table 1 of Ref. \cite{Ade:2013zuv}}
\label{Table:prior-MGmodels}
\end{table}

The cosmological models previously discussed in Section \ref{Section:models} were implemented in our Boltzmann solver EFCLASS. For a given cosmological model and a set of cosmological parameters we can compute the solution for both background and linear perturbations, that is, we can predict observables such as the matter power spectrum and the CMB angular power spectrum. Since the parameter space not only includes cosmological parameters, but also several nuissance parameters, it becomes hard to find the best fit model as well as the relevant statistical information. The usual approach is to use Markov Chain Monte Carlo (MCMC) techniques \cite{Lewis:2002ah,2017ARA&A..55..213S} and we will do so.   

We explore the parameter space of the cosmological models with the code Montepython \cite{Audren:2012wb,Brinckmann:2018cvx} which works along with EFCLASS: theoretical predictions are computed and compared to observations through likelihood functions $\sim 10^5$ times. The MCMC procedure allows us not only to find the best fit model parameters, but also to obtain the relevant countours confidence. In our analysis we use the set of flat priors in Table \ref{Table:prior-MGmodels}.

\subsection{Results}


\subsubsection{DES-fR model}
     
\begin{figure*}[t]
\includegraphics[width=\textwidth]{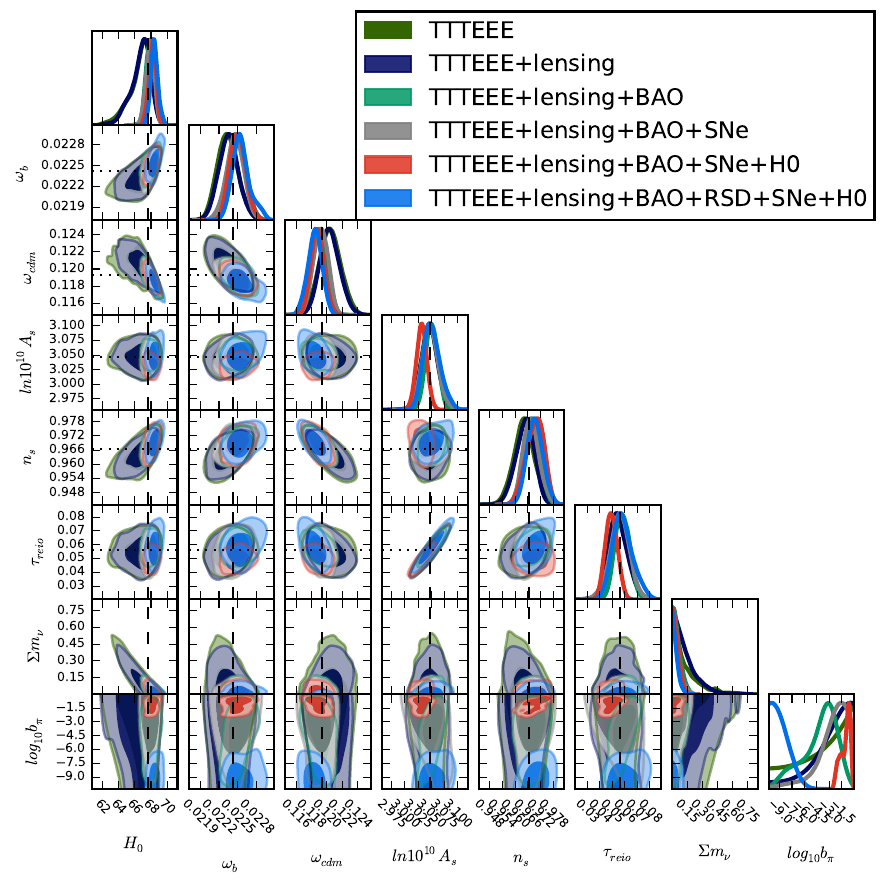}
\caption{1D marginalised likelihoods as well as confidence contours (i.e., $68\%$ and $95\%$) for the DES-fR model. The dashed vertical and horizontal dotted lines correspond to the results obtained by the Planck Collaboration for the \lcdm parameters (last column in Table 2 of Ref. \cite{Aghanim:2018eyx}). $\rm{TTTEEE}$ stands for CMB temperature and E-mode polarisation anisotropies correlations and cross-correlations, $\rm{lensing}$ stands for CMB lensing, $\rm{BAO}$ stands for Baryonic Acoustic Oscillations, $\rm{SNe}$ stands for supernovae, $\rm{H0}$ stands for the Hubble constant, and $\rm{RSD}$ stands for redshift space distortions.}
\label{triangle_DES}
\end{figure*}

In Fig. \ref{triangle_DES} we show the $68\%$ and $95\%$ confidence contours for the DES-fR model. Vertical dashed and horizontal dotted lines indicate the values obtained by the Planck Collaboration in their analyses for the standard cosmological model \lcdm (last column in Table 2 of Ref. \cite{Aghanim:2018eyx}). The relevant statistical information (i.e., mean values and $68\%$ confidence limits) is shown in Table \ref{Table:Designer_fR_Parameters}. We see there is good agreement for common parameters in both DES-fR and \lcdm models. Although error on neutrino masses get significantly reduced as we add more data, we can only set an upper limit when combining all data sets. In the case of the MG  parameter $b_\pi$ we do not observe any degeneracy with other parameters in the model. It is interesting that the constraints on MG, although still prior dominated, present different tendencies according to the combination of data sets: i) RSD push the MG constraints towards GR, while the $H_0$ tension remains unresolved; ii) if we exclude RSD from the data sets, we notice a preference for a MG scenario, but still hitting the prior bound on the right and not solving the problem with $H_0$; iii) a similar situation occurs when we exclude supernovae, $H_0$, and RSD from the data sets, because there is a preference for MG (prior dominated though) while obtaining a $H_0$ value that agrees very well with Planck Collaboration results for \lcdm. Finally, we note that our derived value for the parameter 
\be
\sigma_8 = 0.815^{+0.009}_{-0.007} \qquad (68\%),
\ee
when including the whole data set, agrees very well with the value found by DES Collaboration $\sigma_8 = 0.807^{+0.062}_{-0.041}$ for the \lcdm model \cite{Abbott:2017wau}.

\begin{table*}[!t]
\centering
\resizebox{\columnwidth}{!}{%
\begin{tabular}{c c c c c c c}
\hline
Parameter & Planck  & $\left\lbrace\dots\right\rbrace$+lensing & $\left\lbrace\dots\right\rbrace$+BAO & $\left\lbrace\dots\right\rbrace$+SNe &
$\left\lbrace\dots\right\rbrace$+$H_0$ & $\left\lbrace\dots\right\rbrace$+RSD \\
\hline
$\omega_b$ & $ 0.02234^{-0.00017}_{+0.00015} $ & $0.02235^{-0.00016}_{+0.00015}$ & $0.02252^{-0.00012}_{+0.00013}$ & $0.02246^{-0.00013}_{+0.00014}$ & $0.02250\pm 0.00013$ & $0.02255^{-0.00017}_{+0.00011}$\\
$\omega_{cdm}$ & $0.1204\pm 0.0014$ & $0.1204^{-0.0014}_{+0.0012}$ & $0.1191^{-0.0009}_{+0.0010}$ & $0.1192^{-0.0009}_{+0.0010}$ & $0.1188^{-0.0009}_{+0.0008}$ & $0.1185^{-0.0009}_{+0.0010}$ \\
$H_0$ & $66.64^{-0.78}_{+1.63}$ & $66.70^{-0.80}_{+1.48}$ & $67.96\pm 0.57$ & $67.91^{-0.47}_{+0.55}$ & $68.18^{-0.39}_{+0.49}$ & $68.42^{-0.47}_{+0.41}$ \\
$ \ln 10^{10}A_s$ & $3.047^{-0.016}_{+0.014}$ & $3.047^{-0.016}_{0.015}$ & $3.051\pm 0.011$ & $3.047\pm 0.015$ & $3.032\pm 0.009 $ & $3.051^{-0.018}_{+0.014} $\\
$n_s$ & $0.9643^{-0.0047}_{+0.0049}$ & $0.9646^{-0.0040}_{+0.0047}$& $0.9679\pm 0.0037$ & $0.9676^{-0.0041}_{+0.0033}$ & $ 0.9694^{-0.0038}_{+0.0046}$ & $0.9693^{-0.0039}_{+0.0036}$ \\
$\tau_{reio} $ & $0.0548^{-0.0080}_{+0.0072}$ & $0.0545^{-0.0082}_{+0.0071}$ & $0.0580^{-0.0058}_{+0.0060}$ & $0.0560^{-0.0071}_{+0.0075}$ & $0.0492^{-0.0051}_{+0.0048}$ & $0.0584^{-0.0092}_{0.0068} $ \\
$\Sigma m_\nu $ & $ <0.171$ & $<0.158$ & $<0.064$ & $<0.058$ & $<0.054$ & $<0.043$\\
$\log_{10} b_\pi$ & $ [-10,0] $ & $ [-10,0] $ & $ -4^{-1}_{+3} $ & $ [-10,0] $ & $ -1.1^{-0.6}_{+0.8}$ & $ <-8 $ \\
\hline
\end{tabular}}
\caption{Mean values and $68\%$ confidence limits on cosmological parameters for the DES-fR model. Here $\left\lbrace\dots\right\rbrace$ stands for the inclusion of data from column on the left.}
\label{Table:Designer_fR_Parameters}
\end{table*}


\subsubsection{HS model}

\begin{figure*}[t]
\includegraphics[width=\textwidth]{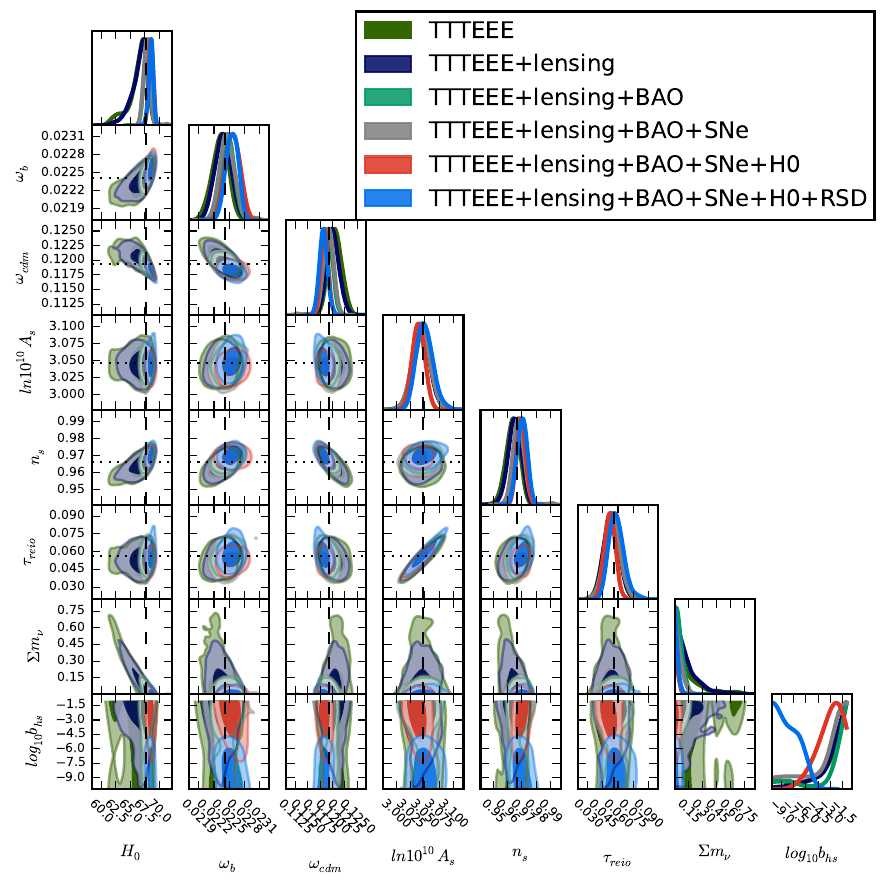}
\caption{1D marginalised likelihoods as well as confidence contours (i.e., $68\%$ and $95\%$) for the HS model. The dashed vertical and horizontal dotted lines correspond to the results obtained by the Planck Collaboration for the \lcdm parameters (last column in Table 2 of Ref. \cite{Aghanim:2018eyx}). $\rm{TTTEEE}$ stands for CMB temperature and E-mode polarisation anisotropies correlations and cross-correlations, $\rm{lensing}$ stands for CMB lensing, $\rm{BAO}$ stands for Baryonic Acoustic Oscillations, $\rm{SNe}$ stands for supernovae, $\rm{H0}$ stands for the Hubble constant, and $\rm{RSD}$ stands for redshift space distortions.}
\label{triangle_HS}
\end{figure*}

In Fig.~\ref{triangle_HS} we depict $68\%$ and $95\%$ confidence contours for the HS model using a number of data sets. Dashed-vertical and dotted-horizontal lines are the parameter values that the Planck Collaboration reported for its analysis using \lcdm model (last column in Table 2 of Ref. \cite{Aghanim:2018eyx}). Table \ref{Table:HS_Parameters} contains relevant statistical information for our analysis: we show mean values and $68\%$ limits for the HS model. Again, cosmological parameters which are common to both \lcdm and HS models are in good agreement with Planck Collaboration's results. As in the case for the DES-fR model we can only find an upper limit for the neutrino masses which is slightly smaller for the HS model. Also in this case MG constraints are prior dominated and we observe a preference for departure from GR in most probe combination, the exception being the case including RSD. The latter again goes towards GR while not solving the $H_0$ discrepancy with the local value. Interestingly, in Ref. \cite{Desmond:2020gzn} the authors analyzed galaxy morphology and placed the following constraint for the HS model $f_{R0}<1.4\times10^{-8}$. By using the whole data set we find  
\be
\sigma_8 = 0.816^{+0.008}_{-0.007} \qquad (68\%),
\ee
which perfectly agrees with the value found for the DES-fR model.

\begin{table*}[!t]
\centering
\resizebox{\columnwidth}{!}{%
\begin{tabular}{c c c c c c c}
\hline
Parameter & Planck  & $\left\lbrace\dots\right\rbrace$+lensing & $\left\lbrace\dots\right\rbrace$+BAO & $\left\lbrace\dots\right\rbrace$+SNe &
$\left\lbrace\dots\right\rbrace$+$H_0$ & $\left\lbrace\dots\right\rbrace$+RSD \\
\hline
$\omega_b$ & $ 0.02233^{-0.00015}_{+0.00018} $ & $0.02238\pm 0.00016 $ & $0.02247\pm 0.00013 $ & $0.02247^{-0.00016}_{+0.00013}$ & $0.02257^{-0.00015}_{+0.00016} $ & $0.02256^{-0.00012}_{+0.00013}$\\
$\omega_{cdm}$ & $0.1205^{-0.0014}_{+0.0013} $ & $0.1201^{-0.0013}_{+0.0012}$ & $0.1191\pm 0.0010 $ & $0.1191^{-0.0009}_{+0.0011}$ & $0.1184^{-0.0008}_{+0.0007}$ & $0.1182\pm 0.0008$ \\
$H_0$ & $66.38^{-0.72}_{+1.89}$ & $66.68^{-0.81}_{+1.70}$ & $67.90^{-0.49}_{+0.59} $ & $67.91^{-0.59}_{+0.52}$ & $68.57^{-0.38}_{+0.40}$ & $68.59^{-0.33}_{+0.43}$ \\
$ \ln 10^{10}A_s$ & $3.046\pm 0.016$ & $3.044\pm 0.014 $ & $3.046\pm 0.014$ & $3.044\pm 0.015$ & $3.039^{-0.010}_{+0.011} $ & $3.050^{-0.016}_{+0.013} $\\
$n_s$ & $0.9638^{-0.0043}_{+0.0051}$ & $0.9650^{-0.0045}_{+0.0047}$& $0.9684^{-0.0040}_{+0.0037} $ & $0.9680^{-0.0043}_{+0.0035}$ & $ 0.9694^{-0.0031}_{+0.0029}$ & $0.9702^{-0.0035}_{+0.0037} $ \\
$\tau_{reio} $ & $0.0535^{-0.0079}_{+0.0077}$ & $0.0536^{-0.0072}_{+0.0074}$ & $0.0557^{-0.0069}_{+0.0073}$ & $0.0544^{-0.0082}_{+0.0073}$ & $0.0519^{-0.0055}_{+0.0053}$ & $0.0584^{-0.0083}_{0.0066} $ \\
$\Sigma m_\nu $ & $ <0.151$ & $<0.143$ & $<0.064$ & $<0.061$ & $<0.026$ & $<0.032$\\
$\log_{10} b_{hs}$ & $ [-10,-1] $ & $>-3 $ & $>-6 $ & $ >-4$ & $ >-4$ & $ <-7 $ \\
\hline
\end{tabular}}
\caption{Mean values and $68\%$ confidence limits on cosmological parameters for the HS model. Here $\left\lbrace\dots\right\rbrace$ stands for the inclusion of data from column on the left.}
\label{Table:HS_Parameters}
\end{table*}

\subsubsection{HDES model}

\begin{figure*}[t]
\includegraphics[width=\textwidth]{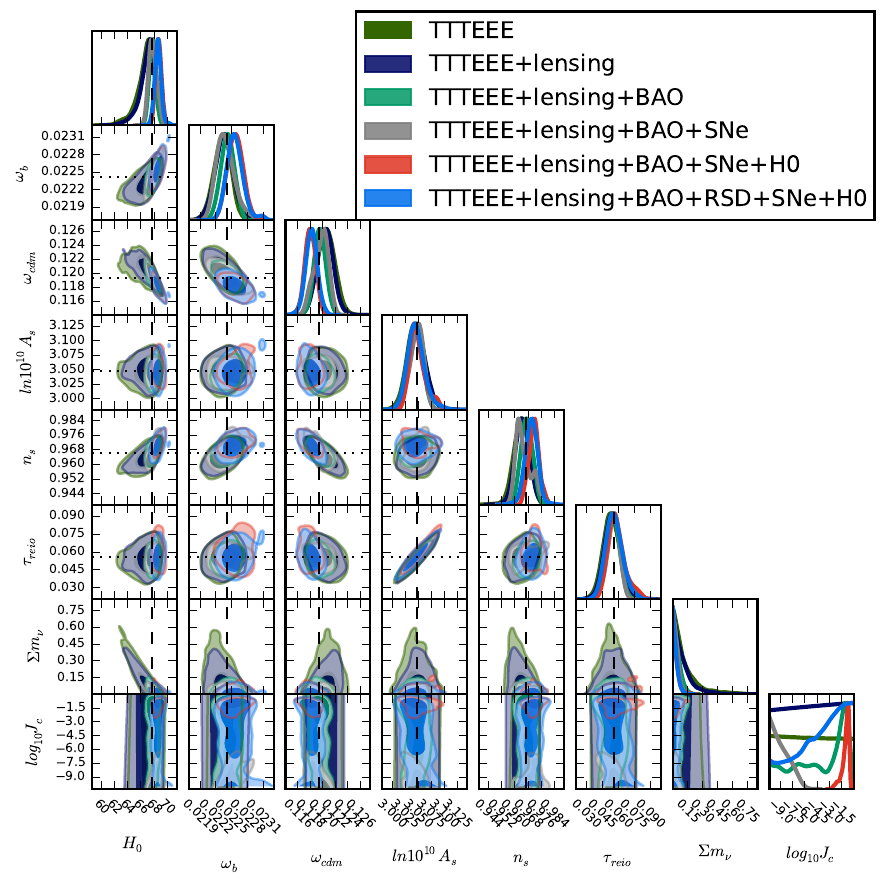}
\caption{1D marginalised likelihoods as well as confidence contours (i.e., $68\%$ and $95\%$) for the HDES model. The dashed vertical and horizontal dotted lines correspond to the results obtained by the Planck Collaboration for the \lcdm parameters (last column in Table 2 of Ref. \cite{Aghanim:2018eyx}). $\rm{TTTEEE}$ stands for CMB temperature and E-mode polarisation anisotropies correlations and cross-correlations, $\rm{lensing}$ stands for CMB lensing, $\rm{BAO}$ stands for Baryonic Acoustic Oscillations, $\rm{SNe}$ stands for supernovae, $\rm{H0}$ stands for the Hubble constant, and $\rm{RSD}$ stands for redshift space distortions.}
\label{triangle_DES_H}
\end{figure*}

Fig. \ref{triangle_DES_H} shows confidence contours for the cosmological parameters in the HDES model. We see good agreement in parameters that also play a part in \lcdm model; the values found by the Planck Collaboration (last column in Table 2 of Ref. \cite{Aghanim:2018eyx}) are depicted as vertical-dashed and horizontal-dotted lines in Fig. \ref{triangle_DES_H}. As for the DES-fR and HS models, in this case the neutrino masses remain unconstrained in our analysis and we can only set an upper limit. Concerning the MG parameter we observe that results are not decisive since posteriors are mostly affected by the prior distribution. Although there exist preference for departure from GR when including $H_0$ and RSD in the data set, the constraints hit the prior bound on the right. Interesting in this case RSD push the constraints far from GR, whereas in the case of DES-fR and HS models the whole data set prefer the GR limit. Finally we note that our derived  
\be
\sigma_8 = 0.814^{+0.009}_{-0.007} \qquad (68\%),
\ee
taking into consideration the full data set agrees well with values found for DES-fR and HS models. 

In Table \ref{Table:Designer_Hordeski_Parameters} we show mean values and $68\%$ confidence bounds for the cosmological parameters in the HDES model.

\begin{table*}[!t]
\centering
\resizebox{\columnwidth}{!}{%
\begin{tabular}{c c c c c c c}
\hline
Parameter & Planck  & $\left\lbrace\dots\right\rbrace$+lensing & $\left\lbrace\dots\right\rbrace$+BAO & $\left\lbrace\dots\right\rbrace$+SNe &
$\left\lbrace\dots\right\rbrace$+$H_0$ & $\left\lbrace\dots\right\rbrace$+RSD \\
\hline
$\omega_b$ & $ 0.02233^{-0.00015}_{+0.00017} $ & $0.02236^{-0.00016}_{+0.00017} $ & $0.02242\pm 0.00013 $ & $0.02238^{-0.00020}_{+0.00014}$ & $0.02257^{-0.00015}_{+0.00014} $ & $0.02257^{-0.00017}_{+0.00013}$\\
$\omega_{cdm}$ & $0.1205\pm 0.0014 $ & $0.1204\pm 0.0013 $ & $0.1195\pm 0.0009 $ & $0.1197^{-0.0010}_{+0.0014}$ & $0.1180\pm 0.0009 $ & $0.1182\pm 0.0009$ \\
$H_0$ & $66.67^{-0.76}_{+1.52}$ & $66.95^{-0.73}_{+1.36}$ & $67.78\pm 0.50 $ & $67.81^{-0.72}_{+0.54}$ & $68.68^{-0.46}_{+0.44}$ & $68.60^{-0.49}_{+0.47}$ \\
$ \ln 10^{10}A_s$ & $3.046^{-0.018}_{+0.016}$ & $3.049^{-0.017}_{+0.015} $ & $3.045\pm 0.013$ & $3.048^{-0.011}_{+0.014}$ & $3.049^{-0.018}_{+0.010} $ & $3.046^{-0.018}_{+0.014} $\\
$n_s$ & $0.9640^{-0.0045}_{+0.0049}$ & $0.9646^{-0.0044}_{+0.0043}$& $0.9671^{-0.0037}_{+0.0035} $ & $0.9658^{-0.0065}_{+0.0069}$ & $ 0.9716^{-0.0032}_{+0.0030}$ & $0.9700^{-0.0037}_{+0.0038} $ \\
$\tau_{reio} $ & $0.0539^{-0.0086}_{+0.0075}$ & $0.0552^{-0.0084}_{+0.0074}$ & $0.0544^{-0.0063}_{+0.0062}$ & $0.0554^{-0.0052}_{+0.0064}$ & $0.0583^{-0.0094}_{+0.0059}$ & $0.0560^{-0.0087}_{0.0056} $ \\
$\Sigma m_\nu $ & $ <0.134$ & $<0.115$ & $<0.056$ & $<0.043$ & $<0.038$ & $<0.037$\\
$\log_{10} J_c$ & $ [-10,0] $ & $ [-10,0] $ & $ <-0.3 $ & $ <-9$ & $ -1.^{-0.2}_{+0.5}$ & $>-5 $ \\
\hline
\end{tabular}}
\caption{Mean values and $68\%$ confidence limits on cosmological parameters for the HDES model. Here $\left\lbrace\dots\right\rbrace$ stands for the inclusion of data from column on the left.}
\label{Table:Designer_Hordeski_Parameters}
\end{table*}

\section{Conclusions\label{Section:conclusions}}

Over the past decades several cosmological models have emerged  as a plausible explanation for the late-time accelerating expansion of the Universe. In this paper we investigated three MG models which satisfy solar system tests and also fulfil constraints on the speed of propagation of GWs, namely: DES-fR, HS, and the HDES models.  

It is possible to interpret MG models as an effective fluid and we followed this approach in this work. We implemented DES-fR, HS, and HDES models in the Boltzmann solver EFCLASS which uses sub-horizon and quasi-static approximations when solving the perturbation equations. We showed in previous works that the observables are accurately computed (i.e., better than $0.1\%$ as compared to outputs from codes which do not use any approximation) while having the advantage of analytical expressions describing MG as an effective fluid. 

When constraining the parameter space for the HS model is usual to assume a \lcdm background. This is however incorrect as the background for the HS model is in general different from the \lcdm model. In this paper we dropped this assumption and solved the perturbations equations taking into consideration the background evolution too. We found constraints which are in good agreement with results by the Planck Collaboration when the parameter spaces overlap. We also note that the constraints on the MG parameter are dominated by the prior hence unconstrained by current data sets. As the HS model has an additional parameter than the \lcdm model, the former will be severely penalized in any Bayesian model comparison. 

Since data indicate a preference for the standard model it is interesting to study models which exactly match the \lcdm background. These models might rely on new physics while also behaving differently at the perturbations level with respect to the \lcdm model. By investigating these kinds of models we can also reveal whether or not current data sets can discriminate alternative models from the concordance model. In this paper we investigated two models, namely, a designer $f(R)$ (DES-fR) and a designer Horndeski model (HDES).

When considering common cosmological parameters, constraints for the DES-fR model do not exhibit significant discrepancies with results by the Planck Collaboration for the standard model. Concerning the MG parameter we note the results depend on the probe combination. Most cases are dominated by the prior and hence unconstrained. The full data set however, prefers the GR limit. One reason for this might be the strong constraints from the RSD likelihood. As the surveys that make the RSD measurements assume a \lcdm model in their analysis, the data themselves maybe a bit biased. While this in general can be corrected, up to a point, with the AP correction as mentioned in the Appendix, some residual bias may remain. While this is an important point, it is however outside the scope of our present analysis, thus we leave it for future work.

Regarding the constraints for the HDES model we also find good agreement with parameters also appearing in the \lcdm model. Here constraints on MG are also inconclusive as the posteriors are prior dominated. Interestingly, in this case the full data set shows a slight preference for a departure from GR. 

In summary, our results do not conclusively indicate the presence of modifications to GR. Since our MG constraints are prior dominated we conclude \lcdm is still the preferred model. Where data sets overlap, our results fully agree with the investigation carried out by the Planck Collaboration \cite{Ade:2015rim}.  

\section*{Acknowledgements}
The authors would like to thank M.~Martinelli for useful discussions. W.C. acknowledges financial support from the Departamento Administrativo de Ciencia, Tecnolog\'{i}a e Innovaci\'{o}n (COLCIENCIAS) under the project ``Discriminaci\'{o}n de modelos de energ\'{i}a oscura y gravedad modificada con futuros datos de cartografiado gal\'{a}ctico" and from Universidad del Valle under the contract 449-2019. S.N. and R.A. acknowledge support from the Research Projects PGC2018-094773-B-C32 and the Centro de Excelencia Severo Ochoa Program SEV-2016-0597. S.N. also acknowledges support from the Ram\'{o}n y Cajal program through Grant No. RYC-2014-15843. The calculations for this article were carried out on the Datacenter CIBioFi. The statistical  analyses  as  well  as  the  plots  were  made with the Python package GetDist \url{https://github.com/cmbant/getdist}.

\section*{Numerical codes}

The numerical codes used by the authors in the analysis of the paper and our modifications to the CLASS code, which we call EFCLASS, can be found on the websites of the EFCLASS \href{https://members.ift.uam-csic.es/savvas.nesseris/efclass.html}{here}, \href{https://github.com/wilmarcardonac/EFCLASS}{here} and \href{https://github.com/RubenArjona/EFCLASS}{here}. The publicly available RSD Montepython likelihood for the growth-rate $f\sigma_8$ data set can be found at \href{https://github.com/snesseris/RSD-growth}{https://github.com/snesseris/RSD-growth}. 

\appendix

\section{The RSD likelihood  \label{Section:appendix-A}}

In this appendix we will present some details on the growth RSD likelihood implemented in MontePython and used to analyse the growth $f\sigma_8$ data of the ``Gold 2018" compilation, namely, the $N=22$ data points of  Ref.~\cite{Sagredo:2018ahx}. This likelihood was first presented in Ref.~\cite{Arjona:2020yum}, but we summarize here again some of the key details for completeness.

The growth data we use in our analysis are based on measurements of the RSD, which is a direct probe of the LSS. In essence, these data directly measure  $f\sigma_8(a)\equiv f(a)\cdot \sigma(a)$, where $f(a) \equiv \frac{d ln\delta}{d lna}$ is the growth rate and $\sigma(a) \equiv \sigma_{8,0}\frac{\delta(a)}{\delta(1)}$ are the redshift-dependent rms fluctuations of the linear density field within spheres of radius $R=8$ Mpc $h^{-1}$. Note that by  $\sigma_{8,0}$ we denote the present value of this parameter. This  dataset has been shown to be internally robust and unbiased by the authors of Ref.~\cite{Sagredo:2018ahx} that used the ``robustness'' criterion of Ref.~\cite{Amendola:2012wc}. The latter uses combinations of subsets in the data to perform a Bayesian analysis and establish the dataset's overall consistency. 

The value of the parameter $f\sigma_8(a)$ can be measured directly by using the ratio of the monopole to the quadrupole of the redshift-space power spectrum. Thus, one can show with linear theory that $f\sigma_8(a)$ does not depend on the bias $b(k,z)$, which maybe both scale and redshift dependent, as it drops from the equations due to the particular combination of variables  \cite{Percival:2008sh,Song:2008qt,Nesseris:2006er}. Furthermore, $f\sigma_8(a)$ has been show to be a good discriminator of DE models \cite{Song:2008qt}. 
The RSD data points are given explicitly in Ref.~\cite{Sagredo:2018ahx} as $ f\sigma_8^\textrm{obs,i}=\Big(f\sigma_8^\textrm{obs}(z_1),\dots, f\sigma_8^\textrm{obs}(z_n)\Big)$. It should be noted that a few of the growth points are correlated, while most also require a fiducial cosmology that has to be adjusted for the Alcock-Paczynski effect (see Refs.~\cite{Sagredo:2018ahx,Nesseris:2017vor,Kazantzidis:2018rnb} as well as Refs.~\cite{Basilakos:2018arq,Basilakos:2017rgc,Basilakos:2016nyg} for earlier analyses). 

The data points that are correlated are the four points from  SDSS \cite{Zhao:2018jxv} and the three WiggleZ points from Ref.~\cite{Blake:2012pj}. Their  covariance matrices are given by
\begin{equation}\label{WiggleZCov}
      \mathbf{C}_{\text{WiggleZ}}= 10^{-3}
    \left(
         \begin{array}{ccc}
           6.400 & 2.570 & 0.000 \\
           2.570 & 3.969 & 2.540 \\
           0.000 & 2.540 & 5.184 \\
         \end{array}
       \right),
\end{equation}
for the WiggleZ data and for the SDSS points by
\begin{equation}\label{SDSS4Cov}
     \mathbf{C}_{\text{SDSS-IV}}= 10^{-2}
    \left(
         \begin{array}{cccc}
   3.098 & 0.892 &  0.329 & -0.021\\
       0.892 & 0.980 & 0.436 & 0.076\\
       0.329 & 0.436 &  0.490   & 0.350 \\
       -0.021 & 0.076 & 0.350 & 1.124
       \end{array}
       \right).
\end{equation}

In order to perform the correction for the Alcock-Paczynski effect, we follow the prescription of Ref.~\cite{Nesseris:2017vor}, which requires the use of a correction factor given by
\begin{equation}
\text{fac}(z^i)= \frac{H(z^i)\,d_A(z^i)}{H^{\text{ref},i}(z^i) \, d_A^{\text{ref},i}(z^i)}\;,
\end{equation}
where the label ``$\text{ref},i$'' stands for the fiducial cosmology at the redshift $z^i$. Then, the corrected data will be given by \cite{Macaulay:2013swa}
\begin{equation}
f\sigma_8^{\textrm{th,i}}\rightarrow\frac{f\sigma_8^{\textrm{th,i}}}{\text{fac}(z^i)}\;.
\end{equation}
By defining the vector $\bm V$ for the data via:
\begin{equation}
     \mathbf{V} =  \mathbf{f\sigma_8^\textrm{obs,i}} -  \frac{f\sigma_8^{\textrm{th,i}}}{\text{fac}(z^i)},
\end{equation}
the chi-squared is then given by 
\begin{equation}
	\chi^2=  \mathbf{V}^T  \mathbf{C}^{-1}  \mathbf{V}\;.
\end{equation}
To conclude, we also need the theoretical prediction for the growth $\delta(k,z)$ at each redshift, which in CLASS can be estimated from the matter power spectrum as $\delta(k,z)=\sqrt{\frac{P(k,z)}{P(k,0)}}$, where we can obtain $P(k,z)$ via the function \texttt{cosmo.pk(k,z)}. Finally, the exact value of the growth-rate $f\sigma_8(k,z)$ can be calculated numerically with direct differentiation and cubic interpolations.

\bibliographystyle{JHEP}
\bibliography{paper}

\end{document}